\def\year{2019}\relax
\documentclass[letterpaper]{article} %
\usepackage{aaai19}  %
\usepackage{times}  %
\usepackage{helvet} %
\usepackage{courier}  %
\usepackage[hyphens]{url}  %
\usepackage{graphicx} %
\usepackage{graphicx}  %
\usepackage{times}

\usepackage{soul}
\usepackage{url}
\usepackage[utf8]{inputenc}
\usepackage{graphicx}
\usepackage{amsmath}
\usepackage{booktabs}
\usepackage{algorithm}
\usepackage{algorithmic}
\usepackage{xspace}
\usepackage{amssymb}
\usepackage{bm}
\usepackage{color}
\usepackage{url}
\usepackage{booktabs}
\usepackage{subfigure}

\newcommand{\OMS}{AMS-A3C\xspace}
\newcommand{\OMF}{AMF-A3C\xspace}

\title{Agent Modeling as Auxiliary Task for Deep Reinforcement Learning}

\author{Pablo Hernandez-Leal,\thanks{Equal contribution} Bilal Kartal$^*$ and Matthew E. Taylor\\
Borealis AI\\
Edmonoton, Canada\\
\{pablo.hernandez, bilal.kartal, matthew.taylor\}@borealisai.com
}

\begin{document}

\maketitle

\begin{abstract}
In this paper we explore how actor-critic methods in deep reinforcement learning, in particular Asynchronous Advantage Actor-Critic (A3C), can be extended with agent modeling. Inspired by recent works on representation learning and multiagent deep reinforcement learning, we propose two architectures to perform agent modeling: the first one based on parameter sharing, and the second one based on agent policy features. Both architectures aim to learn other agents' policies as auxiliary tasks, besides the standard actor (policy) and critic (values). We performed experiments in both cooperative and competitive domains. The former is a problem of coordinated multiagent object transportation and the latter is a two-player mini version of the Pommerman game. Our results show that the proposed architectures stabilize learning and outperform the standard A3C architecture when learning a best response in terms of expected rewards.
\end{abstract}

\section{Introduction}

An important ability for agents to have is to reason about the behaviors of other agents by constructing models that make predictions about the modeled agents~\cite{Albrecht:2018dp}. This \emph{agent modeling}~\cite{schadd2007opponent}\footnote{Sometimes referred as \emph{opponent modelling} since ``opponent" is used to refer to another agent in the environment.} area usually takes concepts and algorithms from multiagent systems (since the environment includes at least two agents), game theory  (which studies the strategic interactions among agents),  and reinforcement learning (since the model may be based on information observed from interactions).

Agent modeling usually serves two purposes in multiagent settings: it improves the coordination efficiency in cooperative scenarios~\cite{Chalkiadakis:2003te} and, in competitive scenarios, it helps the agent to better optimize (best respond) its actions against the predicted opponent policy~\cite{Carmel:1995wh}, e.g., by exploiting opponent mistakes.

Early algorithms for agent modeling came from game theory literature, e.g., fictitious play~\cite{Brown:1951vc}. Later, many works adapted reinforcement learning algorithms for this task~\cite{Banerjee:2005wq}. Recently, agent modeling has been also considered in the context of deep reinforcement learning (DRL).

DRL has shown outstanding results in Atari games, Go, Poker and recently in strategy video games~\cite{Mnih:2013wp,torrado2018deep}. Due to these successes, it is natural that DRL is now being tested in multiagent environments~\cite{hernandez2018multiagent}. Some works have explored using DRL to evaluate emergent behaviors in multiagent environments~\cite{Tampuu:2017fc}, and others have proposed algorithms for multiagent DRL~\cite{Foerster:2017ti}. In contrast, our goal is to estimate other agents' (opponent or teammate) policies by means of an \emph{auxiliary task} at the same time that the agent is learning its respective (best response) policy.
In general, (self-supervised) auxiliary tasks are not used for anything other than shaping the features of the agent, i.e., facilitating the representation learning process~\cite{bellemare2019geometric}, improving learning stability~\cite{Jaderberg:2016uv}, and have broadened the horizons of RL to learn from all experience, whether rewarded or not. Self-supervision defines losses via surrogate annotations that are synthesized from unlabeled inputs. %
Examples are reward prediction which can be cast into a regression task~\cite{Jaderberg:2016uv} and dynamics prediction that captures state, action, and successor relationships. Since the purpose is representation learning and not full modeling of the dynamics and reward, the losses need not form a transition model and proxies can suffice to help tune the representation, i.e., these losses are expected to give gradients not necessarily a generative model~\cite{shelhamer2016loss}.

In this work we take advantage of auxiliary tasks when learning a best response and the opponent/teammate model. Since these two elements are linked to each other, we propose two architectures that take advantage of this realization.

Recently, \emph{asynchronous} actor-critic methods have become widely used in DRL; Asynchronous Advantage Actor-Critic (A3C)~\cite{mnih2016asynchronous} is a major representative of this category, which does not use an experience buffer and learns completely on-policy. Thus, we take A3C as baseline and set off to evaluate and better understand the use of agent modeling as auxiliary task with on-policy actor-critic methods in DRL with the following contributions:

\begin{itemize}
    \item Agent modeling in DRL is still an open research area with opportunities in video games~\cite{zhao2009learning,torrado2018deep,borovikov2019winning}. Our experiments are performed in two recent multiagent environments, one cooperative and one competitive domain (mini version of the Pommerman game).
    \item  We propose two new architectures that take inspiration from multiagent DRL and representation learning to do agent modeling. The first architecture, Agent Modeling by parameter Sharing (\OMS), takes inspiration from the concept of \emph{parameter sharing} to learn the opponent/teammate policy as an \emph{auxiliary task} as well as the standard actor and critic.
    \item The second architecture, Agent Modeling by policy Features (\OMF), leverages the concept of \emph{policy features} to learn latent space features that are used as input when computing the actor and critic of the learning agent.
\end{itemize}

Our results show that modeling the opponent/teammate increases the expected rewards and improves the stability of the learning process. In particular, in this work we show the benefits of using opponent/teammate policy prediction as an auxiliary task with respect to non-learning stochastic agents in both cooperative and competitive scenarios.

\section{Related Work}
In this section, we describe related work on agent modeling in DRL, multiagent DRL and auxiliary tasks.

Deep Reinforcement Opponent Network (DRON)~\cite{He:2016up} was the first DRL work that performed opponent modeling. DRON's idea is to have two networks: one learns $Q$ values (similar to DQN~\cite{Mnih:2013wp}) and a second learns a representation of the opponent's policy. DRON used hand-crafted features to define the opponent network. In contrast, Deep Policy Inference Q-Network (DPIQN) and Deep Policy Inference Recurrent Q-Network (DPIRQN)~\cite{Hong:2018vp} learned opponent \emph{policy features} directly from raw observations of the other agents. The way to learn these policy features is by means of auxiliary tasks~\cite{Jaderberg:2016uv} that provide additional learning goals; in this case, the auxiliary task is to learn the opponent's policy. Then, the $Q$ value function of the learning agent is conditioned on the policy features, which aim to reduce the non-stationarity of the multiagent environment. In contrast, our proposals do not need an experience replay buffer, learn completely on-policy and we make use of full parameter sharing~\cite{Foerster:2017ti}. %

Deep Cognitive Hierarchies~\cite{Lanctot:vn} is an algorithm that aims to avoid overfitting in two-player games. It uses deep reinforcement learning to compute best responses to a distribution over policies and empirical game-theoretic analysis to compute new meta-strategy distributions. Theory of Mind Network~\cite{Rabinowitz:2018uf} tackles the problem of meta-learning, i.e., the proposed network should acquire a strong prior model for agents’ behaviour to bootstrap to richer predictions. DeepBPR+ studies the problem of efficient policy detection and reuse when playing against non-stationary agents in Markov games~\cite{deepbpr2018}. In contrast, our goal is to estimate the opponent/teammate's policy at the same time that the agent is learning its respective (best response) policy; since these two elements are linked to each other our proposals improve the stability of the learning process as well as increase the obtained rewards.

Self Other Modeling (SOM)~\cite{Raileanu:2018tb} is a recently proposed algorithm that uses the agent's own policy as a means to predict the opponent's goal (and actions). SOM is based on the assumption that the agents are identical, which is more suitable when agents share a fixed set of goals and have similar abilities.%

Auxiliary tasks were originally presented as \emph{hints} that improved the network performance and learning time. Suddarth and Kergosien~(\citeyear{suddarth1990rule}) presented a minimal example of a small neural network where it was shown that adding an auxiliary task effectively removed local minima. Recently, some works have used them in single-agent RL problems, for example, Mirowski et al.~(\citeyear{mirowski2016learning}) studied self-supervised tasks (like depth prediction) in a navigation problem. Their results show that augmenting an RL agent with auxiliary tasks supports representation learning, which provides richer training signals that enhance data efficiency. Another interesting result is that using the auxiliary task as a loss was better than using the value as input. Another example was presented by Lample and Chaplot~(\citeyear{lample2017playing}) who added an auxiliary task (to predict game feature information such as the presence of enemies or items) to a DQN network to improve learning in First-Person-Shooting games. These ideas also relate to Multi-Task Learning where by learning tasks in parallel using a shared representation, what is learned for each task can help the learning of the others~\cite{caruana1997multitask}.

\section{Preliminaries}

We start with the standard reinforcement learning setting of an agent interacting in an environment over a discrete number of steps. At time $t$ the agent in state $s_t$ takes an action $a_t$ and receives a reward $r_t$. The state-value function is the expected return (sum of discounted rewards) from state $s$ following a policy $\pi(a|s)$:
$$V^\pi(s)=\mathbb{E}[R_{t:\infty}|s_t=s,\pi],$$
and the action-value function is the expected return following policy $\pi$ after taking action $a$ from state $s$:
$$Q^\pi(s,a)=\mathbb{E}[R_{t:\infty}|s_t=s, a_t=a,\pi].$$

Algorithms, such as Q-learning, or its (deep) neural network variant, DQN, approximate the action-value function $Q(s, a; \theta)$ using parameters $\theta$, and then update parameters to minimize the mean-squared error, using the loss function:
$$
L_Q(\theta_i) = \mathbb{E} [(r + \gamma max_{a'}Q(s', a';\theta_{i}^-) - Q(s, a; \theta_i))^2]
$$
where $\theta^-$ represents the parameters of the target network that is held constant, but synchronized to the behaviour network $\theta^- = \theta$, at certain periods to stabilize learning. %

A3C (Asynchronous Advantage Actor-Critic) is an algorithm that employs a \emph{parallelized} asynchronous training scheme (e.g., using multiple CPUs) for efficiency; it is an on-policy RL method that does not need an experience replay buffer. A3C allows multiple workers to simultaneously interact with the environment and compute gradients locally. All the workers pass their computed local gradients to a global network that performs the optimization and synchronizes the updated actor-critic NN parameters with the workers asynchronously. A3C maintains a parameterized policy (actor) $\pi(a|s;\theta)$ and value function (critic) $V(s; \theta_v)$, which are updated as follows:
$\triangle \theta = \nabla_\theta \log \pi(a_t|s_t; \theta) A(s_t, a_t; \theta_v) $ and  $\triangle \theta_v = A(s_t, a_t; \theta_v) \nabla_{\theta_v} V(s_t)$ where
$$A(s_t, a_t; \theta_v) = \sum_{k=0}^{n-1} \gamma^kr_{t+k} + \gamma^n V(s_{t+n}) - V (s_t),$$ with $A(s,a)=Q(s,a)-V(s)$ representing the \emph{advantage} function, commonly used to reduce variance.

The policy and the value function are updated after every $t_{max}$ actions or when a terminal state is reached. It is common to use one softmax output for the policy head $\pi(a_t|s_t; \theta)$ and one linear output for the value function head $V (s_t; \theta_v)$, with all non-output layers shared, see Figure~\ref{fig:architectures}~(a).

The loss function for A3C is composed of two terms: policy loss (actor), $\mathcal{L}_{\pi}$, and value loss (critic), $\mathcal{L}_{v}$. An entropy loss for the policy, $H(\pi)$, is also commonly added to help to improve exploration by discouraging premature convergence to suboptimal deterministic policies~\cite{mnih2016asynchronous}. Thus, the loss function is given by: $$\mathcal{L}_{\text{A3C}} = \lambda_v  \mathcal{L}_{v} + \lambda_{\pi} \mathcal{L}_{\pi} - \lambda_{H} \mathbb{E}_{s \sim \pi} [H(\pi(s, \cdot, \theta)] $$ with $\lambda_{v}=0.5$, $\lambda_{\pi}=1.0$, and $\lambda_{H}=0.01$, being standard weighting terms on the individual loss components.

The UNsupervised REinforcement and Auxiliary Learning (UNREAL) framework~\cite{Jaderberg:2016uv} is built on top of A3C. UNREAL proposes unsupervised \emph{auxiliary tasks} to speed up the learning process that requires no additional feedback from the environment. The idea of additional auxiliary predictions is to help with the representational learning problem~\cite{bengio2013representation} and had also been useful to improve the robustness and stability of the learning process~\cite{Jaderberg:2016uv}. UNREAL proposes two auxiliary tasks: auxiliary control and auxiliary prediction that share the previous layers that the base agent uses to act. By using this jointly learned representation, the base agent learns to optimize extrinsic reward much faster and, in many cases, achieves better policies at the end of training. The UNREAL algorithm optimizes a single combined loss function with respect to the joint parameters of the agent that combines the A3C loss, $\mathcal{L}_{\text{A3C}}$, together with an auxiliary control loss $\mathcal{L}_{PC}$, an auxiliary reward prediction loss $\mathcal{L}_{RP}$ and a replayed value loss $\mathcal{L}_{VR}$. %
In contrast to A3C, UNREAL uses an experience replay buffer that is sampled with more priority given to interactions with positive rewards to improve the critic network.

\section{Agent Modeling with A3C}

In this section we first describe the challenges of opponent modeling in the context of reinforcement learning and multiagent systems, then we present our two main contributions: the \OMS and \OMF architectures.

\subsection{Opponent modeling and multiagent systems}

In a multiagent environment, agents interact at the same time with the  environment and potentially with each other~\cite{Tuyls:2012up}. These environments are commonly formalized as a Markov game $\langle S, \mathcal{N}, A, T, R \rangle$, which can be seen as an extension of an MDP to multiple agents~\cite{Littman:1994ta}. One key distinction is that the transition, $T$, and reward function, $R$, depend on the actions of all, $\mathcal{N}$, agents.
Given a learning agent $i$ and using the common shorthand notation $\bm{-i} = \mathcal{N} \setminus \{ i \}$ for the set of opponents, the value function now depends on the joint action $\bm{a} = (a_i, \bm{a_{-i}})$, and the joint policy $\bm{\pi}(s, \bm{a}) = \prod_j \pi_j (s, a_j) $:

\begin{equation}
\begin{split}
\label{eqn:bellmanMAS}
V^{\bm{\pi}}_{i}(s)=  \sum_{\bm{a} \in A} \bm{\pi}(s,\bm{a})  \sum_{s' \in S}  T(s,a_i,\bm{a_{-i}},s')\\ [R(s,a_i,\bm{a_{-i}},s') + \gamma V_{i}(s')].
\end{split}
\end{equation}
The optimal policy is dependant on the other agents' policies: %
$$\pi_i^*(s,a_i,\bm{\pi_{-i}}) =  arg\max_{\pi_i} V^{(\pi_i, \bm{\pi_{-i}})}_{i}(s).$$

However, if the other agents' policies are stationary (can still be stochastic) then the problem can be reduced to a standard MDP where RL algorithms can be used to effectively learn a best response to those other agents, irrespective if the domain is cooperative or competitive. Our goal therefore is to accurately estimate the opponent/teammate policy at the same time that the agent is learning its respective (best response) policy. Since these two elements are linked to each other, below we propose two architectures that take advantage of this realization. In this work we show advantage of agent policy prediction with respect to non-learning agents. We leave as future work how to deal with learning agents.

\begin{figure*}
\centering
\includegraphics[scale=0.9]{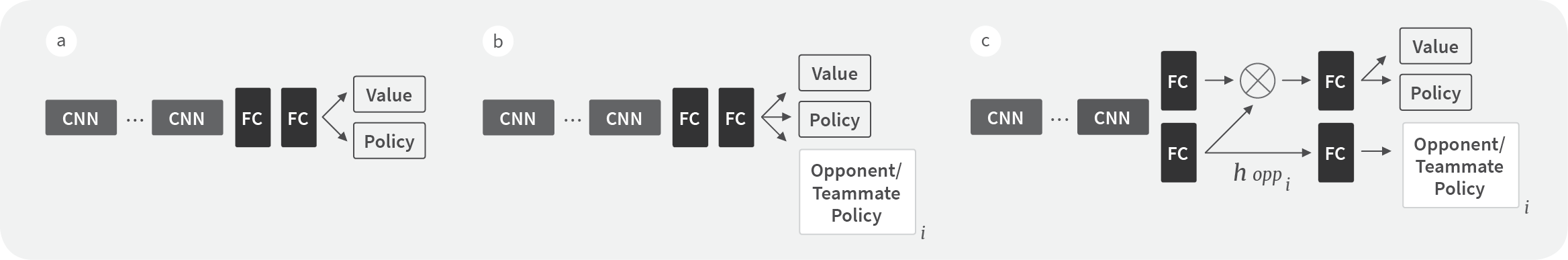}
\caption{CNN represents convolutional layers, FC represents fully connected layers, and $\otimes$ represents an element-wise vector multiplication. (a) A3C outputs values and the agent's policy. (b) \OMS is similar to A3C but adds a head that predicts the other agents' policies. (c) \OMF aims to learn other agents' policy features in the latent space, $h_{opp_i}$, which are then used to compute the value and policy of the learning agent. Both approaches can be generalized to $\mathcal{N}$ opponents/teammates.
}
\label{fig:architectures}
\end{figure*}

\subsection{\OMS: Agent Modeling by parameter Sharing}

This architecture builds on the concepts of \emph{parameter sharing} and \emph{auxiliary tasks}. Parameter sharing has been proposed in multiagent DRL as a way to reduce the number of parameters to learn and improve the performance. The idea is to perform  centralized learning where agents share the same network (i.e., parameters) but the outputs represent different agent actions~\cite{Foerster:2017ti}.

Building on the same principle, in our architecture we want to also predict the opponent/teammate policies as well as the standard actor and critic, with the key characteristic that the previous layers will share all their parameters, see Figure~\ref{fig:architectures}~(b). The change in the architecture is accompanied by a modification in the loss function. In this case, we treat the learning of the other agents' policies as auxiliary tasks~\cite{Jaderberg:2016uv} by refining the loss function as:
$$\mathcal{L}_{\text{\OMS}}= \mathcal{L}_{\text{A3C}} + \frac{1}{\mathcal{N}} \sum_i^{\mathcal{N}} \lambda_{AM_i} \mathcal{L}_{AM_i}$$
where and $\lambda_{AM_i}$ is weight term and $\mathcal{L}_{AM_i}$ is an auxiliary loss for opponent/teammate $i$:
$${\mathcal{L}_{AM_i}= -\frac{1}{M} \sum_j^M a^j_{i} \log (\hat{a}^j_{i}) }$$ %
which is the supervised cross entropy loss between the observed one-hot encoded opponent/teammate action (ground truth), $a^j_{i}$, and the prediction, $\hat{a}^j_{i}$, for a trajectory of length $M$.

\subsection{\OMF: Agent Modeling by policy Features}
The second architecture uses the concepts of \emph{policy features} and auxiliary tasks. Hong et al.~(\citeyear{Hong:2018vp}) proposed a modified DQN architecture that conditions Q-values of the learning agent on \emph{features} in the latent space\footnote{Not to be confused with latent variables.} that also predict the opponent/teammate policy, i.e., policy features.

We take a similar approach but in the context of on-policy actor-critic methods, which means that policy features condition on both actor and critic. In this case, after the convolutional layers, the fully connected layers are divided in two sections, one specialized in the opponent/teammate policy and the other in the actor and critic (of the learning agent). Then, we directly use opponent/teammate policy features, $h_{opp_i}$ vector, to be conditioned (via an element-wise multiplication) when computing the actor and critic, see Figure~\ref{fig:architectures}~(c). The loss function is similarly refined as follows: %

$\mathcal{L}_{\text{\OMF}}= \mathcal{L}_{\text{A3C}} + \frac{1}{\mathcal{N}} \sum_i^{\mathcal{N}} \lambda_{AM_i} \mathcal{L}_{AM_i}$

Note that we described \OMS and \OMF in the general case with $\mathcal{N}$ agents in the environment. In the experiments we evaluated scenarios with one opponent or one teammate.

\subsection{Implementation details}
For A3C, \OMS, and \OMF we used 3 or 4 convolutional layers (depending on the domain), with 32 filters, $3\times3$ kernels, stride and padding of 1. For A3C and \OMS the convolutional layers are followed with 2 fully connected layers with 128 hidden units each, followed by 2-heads: the critic has a single output for state-value estimate for the observation, and the actor has $|A|$ outputs for the policy probabilities for the given observation. For \OMF, the fully connected layers have 64 units (to keep the same number of weights as \OMS). For \OMS and \OMF, the opponent/teammate policy head has $|A_{opp}|$ outputs corresponding to the opponent/teammate policy. We used ELU %
activation functions. The parameters of all architectures have entropy weight of 0.01, a value loss weight of 0.5, a policy loss weight of 1, and a discount factor of 0.99. The parameters of the learning agent's policy are optimized using Adam with $lr=0.0001, \beta_1 = 0.9, \beta_2 = 0.999, \epsilon= 1\times10^{-8}$, and weight decay $1\times10^{-5}$. In the next section we compare different settings for $\lambda_{AM}$.

\section{Experiments}

This section describes the two experimental domains: a cooperative multiagent transport moving problem and a competitive mini version of Pommerman. We then present the experimental results in terms of sensitivity of the loss weight parameter $\lambda_{AM}$ for \OMS and \OMF in the coordination domain, and then we compare with A3C in terms of rewards for the two domains.

\subsection{Domains and setup}

\begin{figure}
\centering
\subfigure[]{
\includegraphics[scale=0.43]{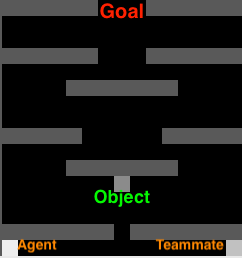}
\label{fig:Boxes}
}
\subfigure[]{
\includegraphics[scale=0.28]{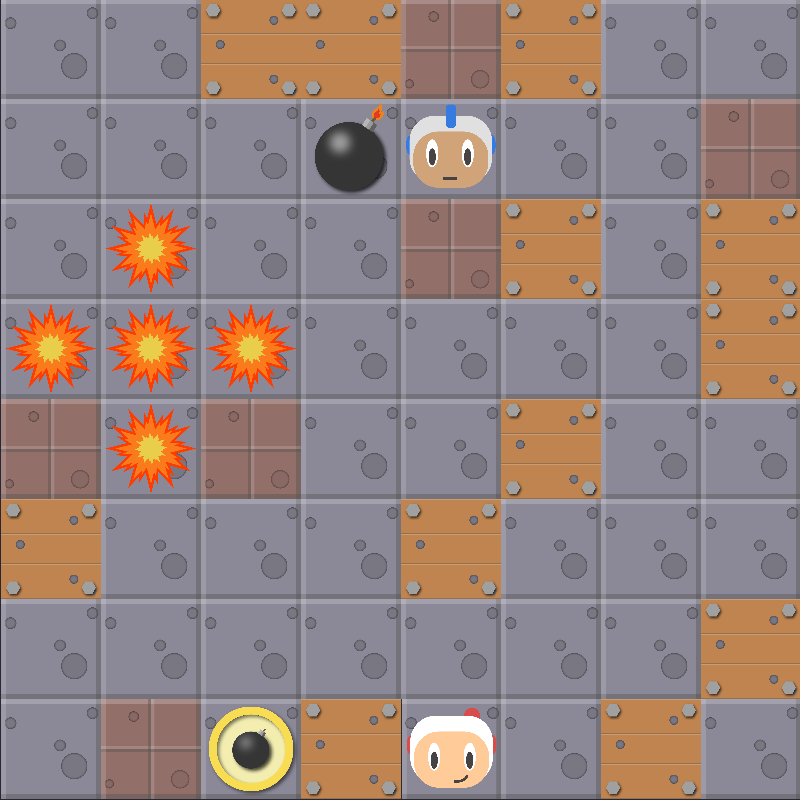}
\label{fig:pom8x8}
}
\caption{(a) The coordinated multiagent object transport moving problem~\protect\cite{Palmer:2018wv}. Two agents need to coordinate to pick up an object and delivery it to the goal zone. Our experiments use a stochastic teammate that moves with higher probability towards the object and then to the goal. (b) An example of the  mini Pommerman with board size 8$\times$8. The board is randomly generated varying the number of wood (light brown cell), walls (dark brown cell) and power-ups (yellow circle). Initial positions of the agents are randomized close to any of the 4 corners of the board.}
\end{figure}

\paragraph{Coordination}
This domain is inspired by Coordinated Multi-Agent Object Transportation Problems (CMOTPs)~\cite{Palmer:2018wv}, in which two agents are tasked with delivering one object to a goal within a grid-world. The agents must locate and pick up the object by standing in the grid cells on the left and right hand side. The task is fully cooperative, i.e., objects can only be transported upon both agents grasping the item (this happens automatically when situated next to the object) and choosing to move in the same direction. Agents only receive a positive reward after placing the object in the goal, see Figure~\ref{fig:Boxes}. Agents have 1900 time steps to complete this task, otherwise the object is reset to the starting position. The actions available to each agent are to either stay in place or move left, right, up, or down. We tested two teammates: one \emph{hesitant} agent which moves randomly but with higher probability towards the object and once it has grasped it then moves with higher probability towards the goal; and a \emph{stubborn} agent which prefers to follow a certain path after grasping the object (i.e., some actions are fully deterministic). Agents receive one observation per time step from the environment as a $16\times16$ pixel representation. We used 12 CPU workers in these experiments. %

\paragraph{Competition}
The Pommerman environment~\cite{resnick2018pommerman}
is based on the classic console game Bomberman. Our experiments use the simulator in a mode with two agents, see Figure~\ref{fig:pom8x8}. Each agent can execute one of 6 actions at every time step: move in any of four directions, stay put, or place a bomb. Each cell on the board can be a passage, a rigid wall, or wood. The maps are generated randomly, albeit there is always a guaranteed path between any two agents. The winner of the game is the last agent standing and receives a reward of 1. Whenever an agent places a bomb it explodes after 10 time steps, producing flames that have a lifetime of 2 time steps. Flames destroy wood and kill any agents within their blast radius. When wood is destroyed either a passage or a power-up is revealed. Power-ups can be of three types: increase the blast radius of bombs, increase the number of bombs the agent can place, or give the ability to kick bombs. A single game of two-player Pommerman is finished when an agent dies or when reaching 800 timesteps.

We considered %
the \emph{rule-based} opponent baseline that comes within the simulator (a.k.a. SimpleAgent). Its behaviour is stochastic since it collects power-ups and places bombs when it is near an opponent. It is skilled in avoiding blasts from bombs. It uses Dijkstra's algorithm on each time-step, resulting in longer training times.

We evaluated our two proposed architectures and the standard A3C against the opponents mentioned above. In all cases we provided learning agents with dense rewards and  we did not tune those reward terms. In our setting the entire board is visible and agents receive one observation per time step from the environment as a $18\times8\times8$ matrix which contains the current time step board description of the board for the current time step, similar to Resnick et al.~(\citeyear{resnick2018backplay}).%

\subsection{Results}
\begin{figure}
\centering
\subfigure[]{
\includegraphics[scale=0.24, clip=true]{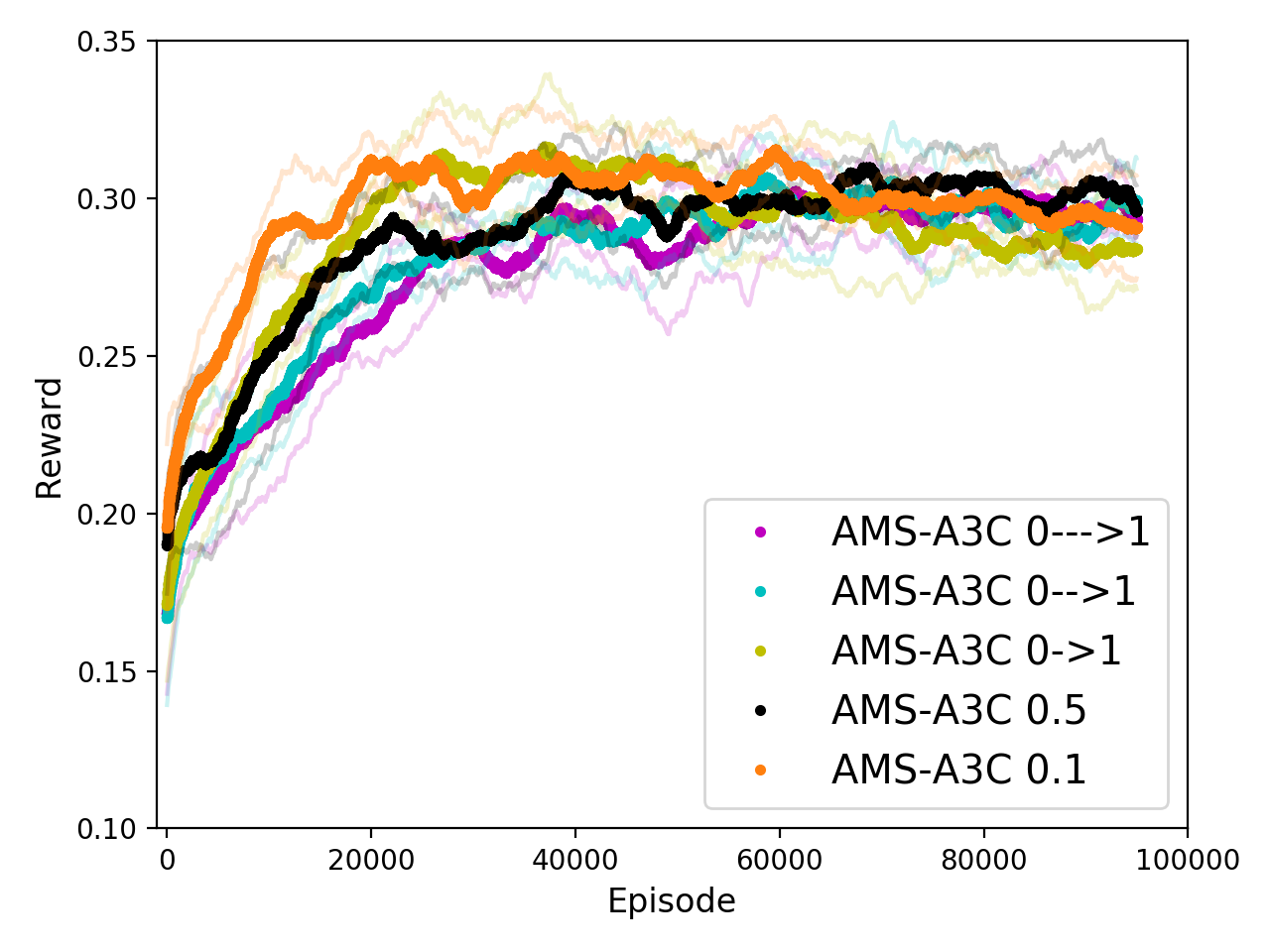}
}
\subfigure[]{
\includegraphics[scale=0.24, clip=true]{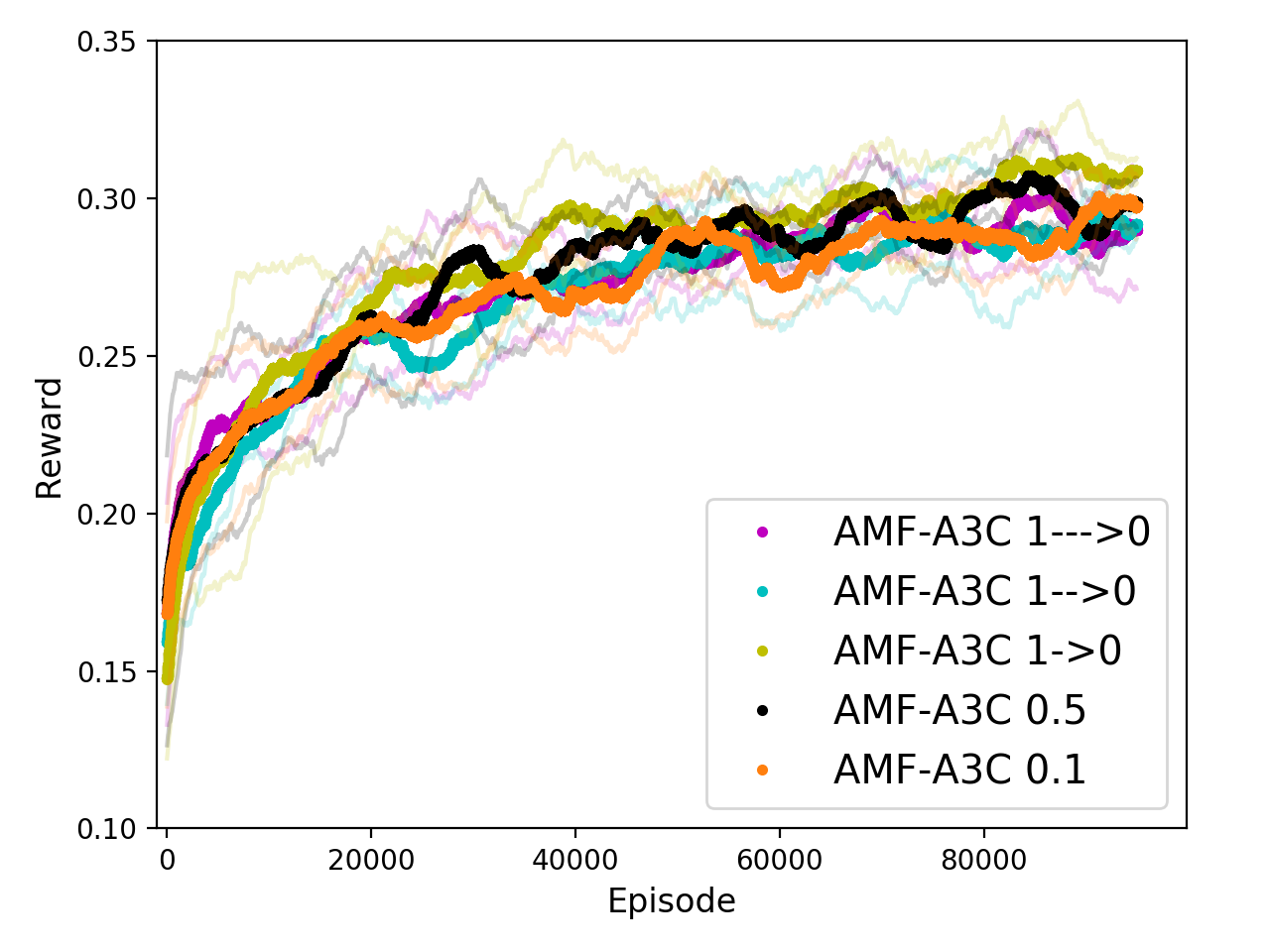}
}
\subfigure[]{
\includegraphics[scale=0.24]{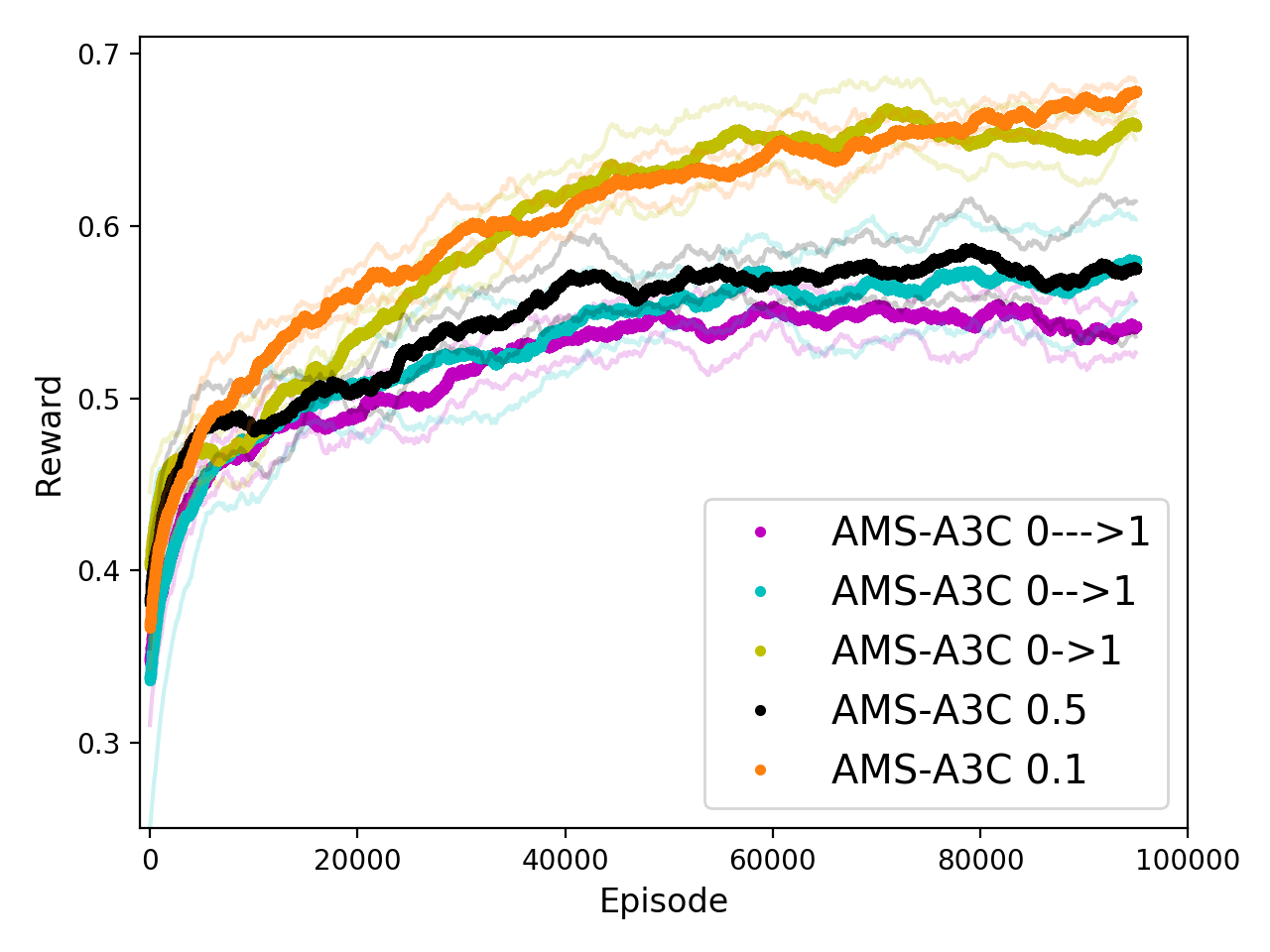}
}
\subfigure[]{
\includegraphics[scale=0.24]{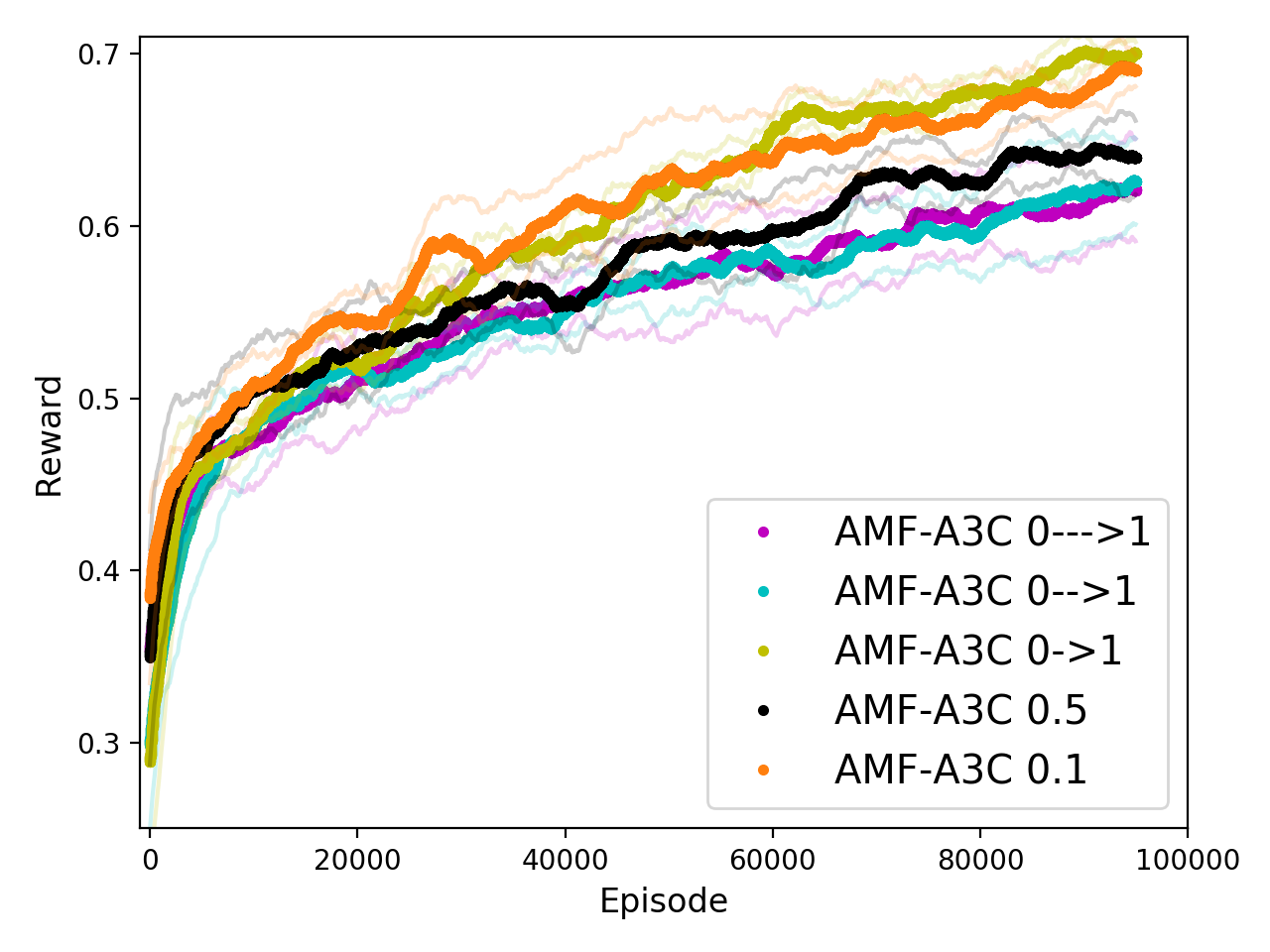}
}
\caption{Comparison for the weight for the opponent modeling loss value, $\lambda_{AM}$, annealing $1.0\rightarrow 0.0$ with varying discount rates (exponentially) or fixing the value. Learning curves in the coordination domain with the \emph{hesitant} teammate for (a) \OMS and (b) \OMF: no significant variation; with the \emph{stubborn} teammate for (c) \OMS  and (d) \OMF:  best results were obtained with $\lambda_{AM} =0.1$}
\label{fig:paramters}
\end{figure}

\paragraph{Sensitivity of $\lambda_{AM}$}

In the first set of experiments we used the coordination domain to evaluate different weights for the opponent modeling loss value: annealing $\lambda_{AM}=1.0 \rightarrow 0.0$ varying discount rates exponentially $\{0.999, 0.9999, 0.99999\}$ or keeping the value fixed with $\lambda_{AM} =\{0.1, 0.5\}$. With the \emph{hesitant} teammate both \OMS and \OMF show similar behavior for all the evaluated parameters (better than A3C), see Figures~\ref{fig:paramters}(a)-(b). When testing with the \emph{stubborn} teammate we observed more variation among parameters, for both \OMS and \OMS using a fixed $\lambda_{AM} =0.1$ or quickly annealing with $0.999$ gave the best results, see Figures~\ref{fig:paramters}(c)-(d). Our hypothesis is that this teammate is easier to learn and the network does not need too much weight on their modeling; instead it can focus on policy learning.

\begin{figure}
\centering
\subfigure[]{
\includegraphics[scale=0.24]{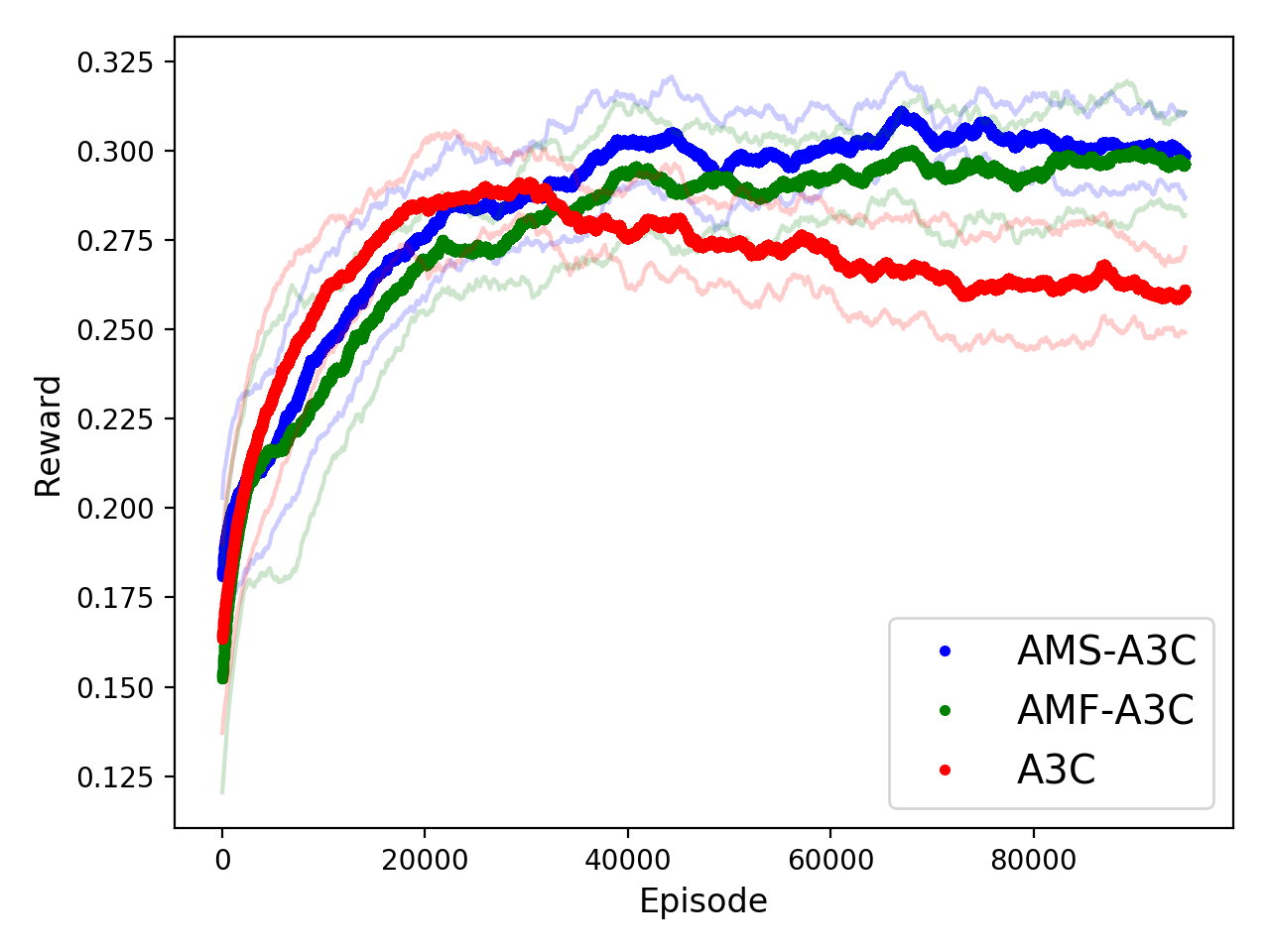}
\label{fig:learningboxesHard}
}
\centering
\subfigure[]{
\includegraphics[scale=0.24]{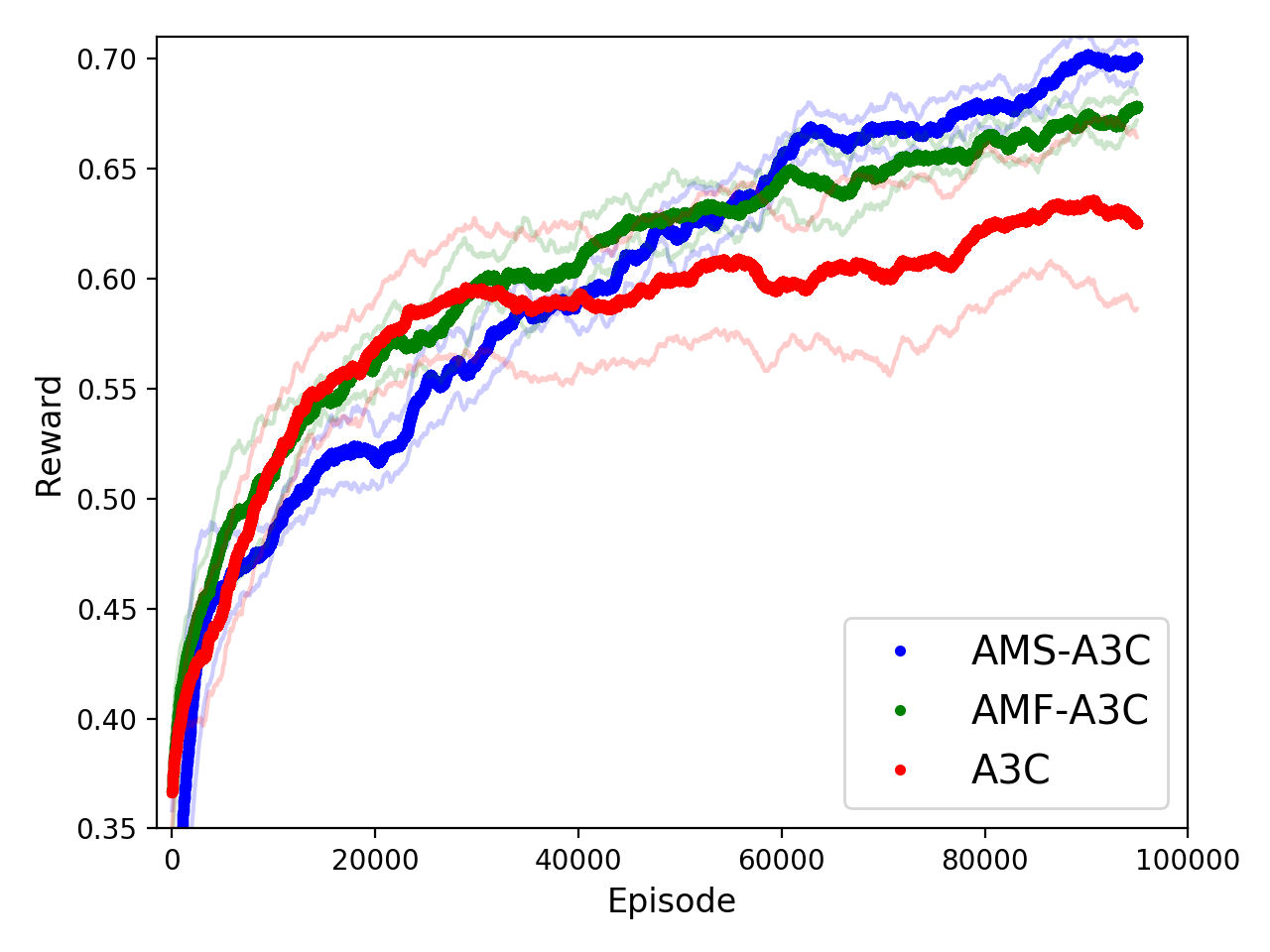}
\label{fig:learningboxesEasy}
}
\caption{Coordination domain: Learning curves with two different teammates (a) \emph{hesitant} and (b) \emph{stubborn} in the coordination problem. Vanilla A3C shows instability and even reduces its rewards after some episodes, in contrast, \OMS and \OMF are more stable, with lower variance and higher rewards.
}
\label{fig:learningboxes}
\end{figure}

\paragraph{Coordination}

Using the best parameters for \OMS and \OMF we compare to A3C. Figure~\ref{fig:learningboxes} depicts learning curves\footnote{Because of the stochasticity of the opponent actions an upper bound of the expected reward is $\approx 0.7$ (experimentally computed) with the selected parameters.} (average with standard deviations of 10 runs) where it can be seen that in the first part of the learning (30k episodes), all learning agents behave similarly, however, in the long run \OMS and \OMF obtained higher rewards than A3C (\OMS was statistically significantly better than A3C from episode 60k, $\alpha=0.05$). We noted that against the \emph{hesitant} teammate A3C decreases its rewards, likely because of its on-policy nature, see Figure~\ref{fig:learningboxesHard}. In contrast, \OMS and \OMF show stability and start increasing their rewards. When facing the \emph{stubborn} teammate, \OMS and \OMF show less variance than A3C due to their accurate agent modeling (\OMS is statistically significant over A3C from episode 90k with $\alpha=0.05$), see Figure~\ref{fig:learningboxesEasy}. Examining the trained agents, \OMS and \OMF show better coordination skills once the object is grasped compared to vanilla A3C, i.e., agents reached the goal faster once grabbing the object.

\paragraph{Competition}

\begin{figure}
\centering
\includegraphics[scale=0.38]{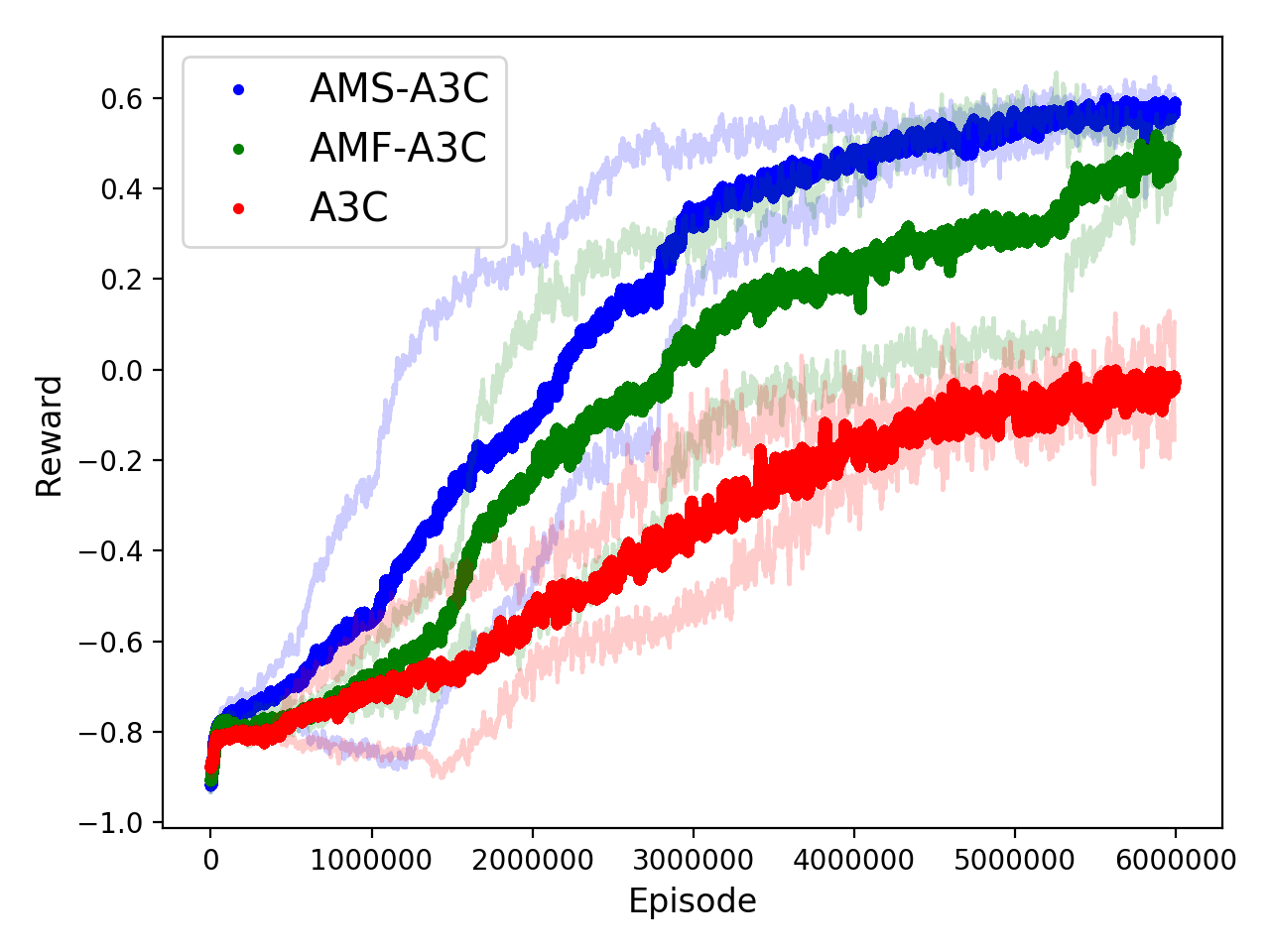}
\caption{Competition domain: Moving average over 10k games of the rewards (shaded lines are standard deviation over 5 runs) obtained by the two proposed architectures and A3C against the \emph{rule-based} opponent: \OMS and \OMF obtained significantly higher scores than A3C.
}
\label{fig:rndSimple}
\end{figure}

\begin{figure}
\centering
\subfigure[]{
\includegraphics[scale=0.31]{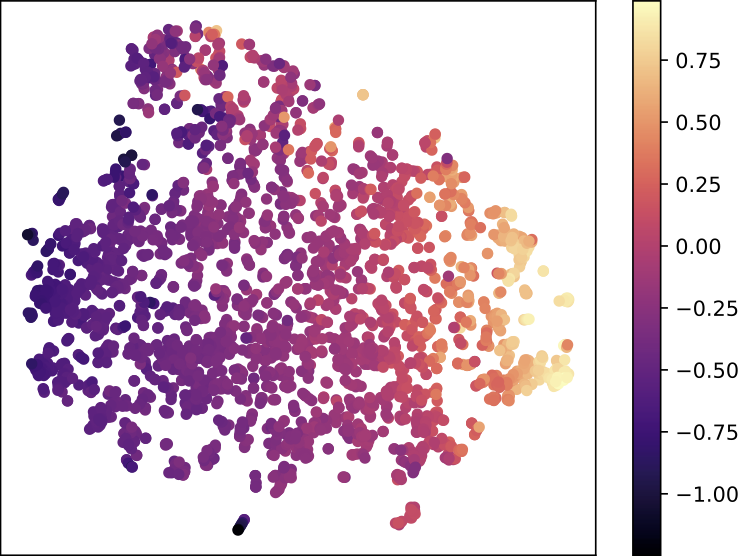}
\label{fig:tsn1}
}
\subfigure[]{
\includegraphics[scale=0.31]{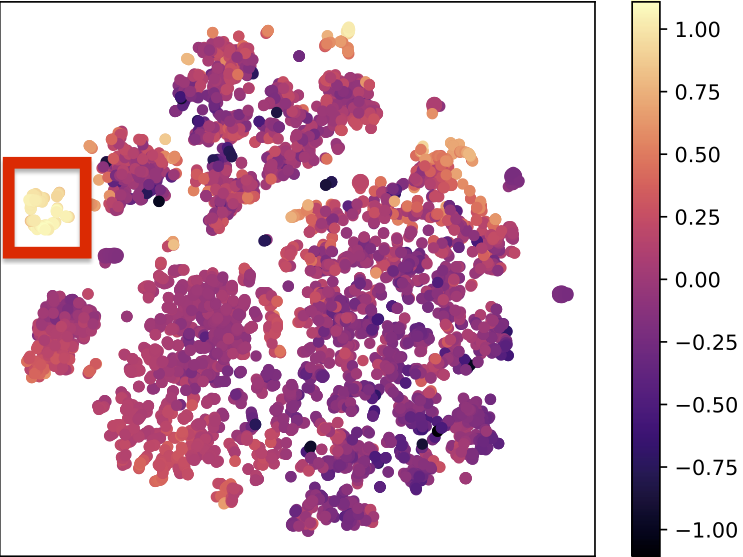}
\label{fig:tsne1}
}
\caption{T-SNE analysis from trained (a) A3C and (b) \OMS agents from 100 episodes (colors are obtained from the value head). \OMS t-SNE shows many more clusters, in particular, the cluster highlighted on the left corresponds to states when \OMS is about to win the game (value close to 1).
}
\label{fig:tsnes}
\end{figure}

One clear distinction from the previous domain is that it is more elaborate and stochastic (board is randomized and changes depending on the agents' actions). In this experiments we set $\lambda_{AM} =0.01$ and we evaluate against the \emph{rule-based} opponent. In this case, we let the learning agents train for 6 million episodes to guarantee convergence ($\approx$ 3 days of training with 50 workers). Results are depicted in Figure~\ref{fig:rndSimple} (with standard deviations over 5 runs), where it can be seen that \OMS and \OMF both clearly outperform A3C in terms of rewards (\OMS is statistically significant over A3C from episode 3.5M and \OMF from 5.5M, $\alpha=0.05$). When observing the policies generated we noted that during game play the agents trained with \OMS and \OMF were able to make the opponent commit suicide by %
blocking its moves (in Pommerman, if two agents simultaneously want to move to the same cell, they both stay in their current locations) and make it stand on the path of the flames, in contrast to A3C which was unable to learn this strategy and obtained lower rewards.

Lastly, we performed a visual analysis similar to Zahavy, Ben-Zrihem and Mannor~(\citeyear{zahavy2016graying}). We took trained agents of A3C and \OMS and for 100 episodes we recorded both the activations of the last layer and the value output. We applied t-SNE ~\cite{maaten2008visualizing} on the activations data (as input) and the value outputs (as labels). Figure~\ref{fig:tsnes} depicts the t-SNEs where it can be seen that \OMS has more well-defined clusters than A3C's, in particular the highlighted cluster on the left represents states when \OMS is about to win the game because it can accurately predict the opponent's moves, which implies values close to 1.

\section{Conclusions}

Deep reinforcement learning has shown outstanding results in recent years. However, there are still many open questions regarding different recent learning algorithms. We take as base a major representative of actor-critic methods, i.e., A3C, and propose two architectures that are designed to do agent modelling as an auxiliary task. This means that at the same time the network improves the representation learning, it will also aim to learn other agents policies. Even though auxiliary tasks are not new, their use in deep RL and opponent modeling is still not well studied. Our work serves as an important stepping stone in this direction by proposing two architectures that improve learning when doing opponent/teammate modeling in deep RL. Our architectures \OMS and \OMF are inspired by multiagent DRL concepts: parameter sharing and opponent policy features. We experimented in both cooperative and competitive domains. In the former, our proposals were able to learn coordination faster and more robustly compared to the vanilla A3C. In the latter, our agents were able to predict opponent moves in complex simultaneous move, Pommerman, and successfully obtain a best response that resulted in higher scores in terms of rewards. As future work, we are interested in exploring self-play, learning agents, and mixed (coordination-competition) environments.

\bibliography{ref}

\begin{thebibliography}{}

\bibitem[\protect\citeauthoryear{Albrecht and Stone}{2018}]{Albrecht:2018dp}
Albrecht, S.~V., and Stone, P.
\newblock 2018.
\newblock {Autonomous agents modelling other agents: A comprehensive survey and
  open problems}.
\newblock {\em Artificial Intelligence} 258:66--95.

\bibitem[\protect\citeauthoryear{Banerjee and Peng}{2005}]{Banerjee:2005wq}
Banerjee, B., and Peng, J.
\newblock 2005.
\newblock {Efficient learning of multi-step best response}.
\newblock In {\em AAMAS},  60--66.

\bibitem[\protect\citeauthoryear{Bellemare \bgroup et al\mbox.\egroup
  }{2019}]{bellemare2019geometric}
Bellemare, M.~G.; Dabney, W.; Dadashi, R.; Taiga, A.~A.; Castro, P.~S.; Roux,
  N.~L.; Schuurmans, D.; Lattimore, T.; and Lyle, C.
\newblock 2019.
\newblock A geometric perspective on optimal representations for reinforcement
  learning.
\newblock {\em arXiv preprint arXiv:1901.11530}.

\bibitem[\protect\citeauthoryear{Bengio, Courville, and
  Vincent}{2013}]{bengio2013representation}
Bengio, Y.; Courville, A.; and Vincent, P.
\newblock 2013.
\newblock Representation learning: A review and new perspectives.
\newblock {\em IEEE transactions on pattern analysis and machine intelligence}
  35(8):1798--1828.

\bibitem[\protect\citeauthoryear{Borovikov \bgroup et al\mbox.\egroup
  }{2019}]{borovikov2019winning}
Borovikov, I.; Zhao, Y.; Beirami, A.; Harder, J.; Kolen, J.; Pestrak, J.;
  Pinto, J.; Pourabolghasem, R.; Chaput, H.; Sardari, M.; et~al.
\newblock 2019.
\newblock Winning isn’t everything: Training agents to playtest modern games.
\newblock In {\em AAAI Workshop on Reinforcement Learning in Games}.

\bibitem[\protect\citeauthoryear{Brown}{1951}]{Brown:1951vc}
Brown, G.~W.
\newblock 1951.
\newblock {Iterative solution of games by fictitious play}.
\newblock {\em Activity analysis of production and allocation} 13(1):374--376.

\bibitem[\protect\citeauthoryear{Carmel and Markovitch}{1995}]{Carmel:1995wh}
Carmel, D., and Markovitch, S.
\newblock 1995.
\newblock {Opponent Modeling in Multi-Agent Systems}.
\newblock In {\em IJCAI}.
\newblock ~Springer-Verlag.

\bibitem[\protect\citeauthoryear{Caruana}{1997}]{caruana1997multitask}
Caruana, R.
\newblock 1997.
\newblock Multitask learning.
\newblock {\em Machine learning} 28(1):41--75.

\bibitem[\protect\citeauthoryear{Chalkiadakis and
  Boutilier}{2003}]{Chalkiadakis:2003te}
Chalkiadakis, G., and Boutilier, C.
\newblock 2003.
\newblock {Coordination in Multiagent Reinforcement Learning: A Bayesian
  Approach}.
\newblock In {\em AAMAS}.

\bibitem[\protect\citeauthoryear{Foerster \bgroup et al\mbox.\egroup
  }{2017}]{Foerster:2017ti}
Foerster, J.~N.; Nardelli, N.; Farquhar, G.; Afouras, T.; Torr, P. H.~S.;
  Kohli, P.; and Whiteson, S.
\newblock 2017.
\newblock {Stabilising Experience Replay for Deep Multi-Agent Reinforcement
  Learning.}
\newblock In {\em ICML}.

\bibitem[\protect\citeauthoryear{He \bgroup et al\mbox.\egroup
  }{2016}]{He:2016up}
He, H.; Boyd-Graber, J.; Kwok, K.; and Daume, H.
\newblock 2016.
\newblock {Opponent modeling in deep reinforcement learning}.
\newblock In {\em ICML},  2675--2684.

\bibitem[\protect\citeauthoryear{Hernandez-Leal, Kartal, and
  Taylor}{2018}]{hernandez2018multiagent}
Hernandez-Leal, P.; Kartal, B.; and Taylor, M.~E.
\newblock 2018.
\newblock {Is multiagent deep reinforcement learning the answer or the
  question? A brief survey}.
\newblock {\em arXiv preprint arXiv:1810.05587}.

\bibitem[\protect\citeauthoryear{Hong \bgroup et al\mbox.\egroup
  }{2018}]{Hong:2018vp}
Hong, Z.-W.; Su, S.-Y.; Shann, T.-Y.; Chang, Y.-H.; and Lee, C.-Y.
\newblock 2018.
\newblock {A Deep Policy Inference Q-Network for Multi-Agent Systems}.
\newblock In {\em AAMAS}.

\bibitem[\protect\citeauthoryear{Jaderberg \bgroup et al\mbox.\egroup
  }{2017}]{Jaderberg:2016uv}
Jaderberg, M.; Mnih, V.; Czarnecki, W.~M.; Schaul, T.; Leibo, J.~Z.; Silver,
  D.; and Kavukcuoglu, K.
\newblock 2017.
\newblock {Reinforcement Learning with Unsupervised Auxiliary Tasks.}
\newblock In {\em ICLR}.

\bibitem[\protect\citeauthoryear{Lample and Chaplot}{2017}]{lample2017playing}
Lample, G., and Chaplot, D.~S.
\newblock 2017.
\newblock {Playing FPS Games with Deep Reinforcement Learning}.
\newblock In {\em AAAI},  2140--2146.

\bibitem[\protect\citeauthoryear{Lanctot \bgroup et al\mbox.\egroup
  }{2017}]{Lanctot:vn}
Lanctot, M.; Zambaldi, V.~F.; Gruslys, A.; Lazaridou, A.; Tuyls, K.;
  P{\'e}rolat, J.; Silver, D.; and Graepel, T.
\newblock 2017.
\newblock {A Unified Game-Theoretic Approach to Multiagent Reinforcement
  Learning.}
\newblock In {\em NIPS}.

\bibitem[\protect\citeauthoryear{Littman}{1994}]{Littman:1994ta}
Littman, M.~L.
\newblock 1994.
\newblock {Markov games as a framework for multi-agent reinforcement learning}.
\newblock In {\em ICML},  157--163.

\bibitem[\protect\citeauthoryear{Maaten and
  Hinton}{2008}]{maaten2008visualizing}
Maaten, L. v.~d., and Hinton, G.
\newblock 2008.
\newblock {Visualizing data using t-SNE}.
\newblock {\em Journal of Machine Learning Research} 9(Nov).

\bibitem[\protect\citeauthoryear{Mirowski \bgroup et al\mbox.\egroup
  }{2017}]{mirowski2016learning}
Mirowski, P.; Pascanu, R.; Viola, F.; Soyer, H.; Ballard, A.~J.; Banino, A.;
  Denil, M.; Goroshin, R.; Sifre, L.; Kavukcuoglu, K.; et~al.
\newblock 2017.
\newblock Learning to navigate in complex environments.
\newblock {\em ICLR}.

\bibitem[\protect\citeauthoryear{Mnih \bgroup et al\mbox.\egroup
  }{2013}]{Mnih:2013wp}
Mnih, V.; Kavukcuoglu, K.; Silver, D.; Graves, A.; Antonoglou, I.; Wierstra,
  D.; and Riedmiller, M.
\newblock 2013.
\newblock {Playing Atari with Deep Reinforcement Learning}.
\newblock {\em arXiv preprint arXiv:1312.5602v1}.

\bibitem[\protect\citeauthoryear{Mnih \bgroup et al\mbox.\egroup
  }{2016}]{mnih2016asynchronous}
Mnih, V.; Badia, A.~P.; Mirza, M.; Graves, A.; Lillicrap, T.; Harley, T.;
  Silver, D.; and Kavukcuoglu, K.
\newblock 2016.
\newblock Asynchronous methods for deep reinforcement learning.
\newblock In {\em ICML},  1928--1937.

\bibitem[\protect\citeauthoryear{Palmer \bgroup et al\mbox.\egroup
  }{2018}]{Palmer:2018wv}
Palmer, G.; Tuyls, K.; Bloembergen, D.; and Savani, R.
\newblock 2018.
\newblock {Lenient Multi-Agent Deep Reinforcement Learning.}
\newblock In {\em AAMAS}.

\bibitem[\protect\citeauthoryear{Rabinowitz \bgroup et al\mbox.\egroup
  }{2018}]{Rabinowitz:2018uf}
Rabinowitz, N.~C.; Perbet, F.; Song, H.~F.; Zhang, C.; Eslami, S. M.~A.; and
  Botvinick, M.
\newblock 2018.
\newblock {Machine Theory of Mind.}
\newblock In {\em ICML}.

\bibitem[\protect\citeauthoryear{Raileanu \bgroup et al\mbox.\egroup
  }{2018}]{Raileanu:2018tb}
Raileanu, R.; Denton, E.; Szlam, A.; and Fergus, R.
\newblock 2018.
\newblock {Modeling Others using Oneself in Multi-Agent Reinforcement
  Learning.}
\newblock In {\em ICML}.

\bibitem[\protect\citeauthoryear{Resnick \bgroup et al\mbox.\egroup
  }{2018}]{resnick2018pommerman}
Resnick, C.; Eldridge, W.; Ha, D.; Britz, D.; Foerster, J.; Togelius, J.; Cho,
  K.; and Bruna, J.
\newblock 2018.
\newblock Pommerman: A multi-agent playground.
\newblock {\em AIIDE Multi-Agent Workshop}.

\bibitem[\protect\citeauthoryear{Resnick \bgroup et al\mbox.\egroup
  }{2019}]{resnick2018backplay}
Resnick, C.; Raileanu, R.; Kapoor, S.; Peysakhovich, A.; Cho, K.; and Bruna, J.
\newblock 2019.
\newblock Backplay:" man muss immer umkehren".
\newblock {\em AAAI-19 Workshop on Reinforcement Learning in Games}.

\bibitem[\protect\citeauthoryear{Schadd, Bakkes, and
  Spronck}{2007}]{schadd2007opponent}
Schadd, F.; Bakkes, S.; and Spronck, P.
\newblock 2007.
\newblock Opponent modeling in real-time strategy games.
\newblock In {\em GAMEON},  61--70.

\bibitem[\protect\citeauthoryear{Shelhamer \bgroup et al\mbox.\egroup
  }{2017}]{shelhamer2016loss}
Shelhamer, E.; Mahmoudieh, P.; Argus, M.; and Darrell, T.
\newblock 2017.
\newblock Loss is its own reward: Self-supervision for reinforcement learning.
\newblock {\em ICLR workshops}.

\bibitem[\protect\citeauthoryear{Suddarth and
  Kergosien}{1990}]{suddarth1990rule}
Suddarth, S.~C., and Kergosien, Y.
\newblock 1990.
\newblock Rule-injection hints as a means of improving network performance and
  learning time.
\newblock In {\em Neural Networks}. Springer.
\newblock  120--129.

\bibitem[\protect\citeauthoryear{Tampuu \bgroup et al\mbox.\egroup
  }{2017}]{Tampuu:2017fc}
Tampuu, A.; Matiisen, T.; Kodelja, D.; Kuzovkin, I.; Korjus, K.; Aru, J.; Aru,
  J.; and Vicente, R.
\newblock 2017.
\newblock {Multiagent cooperation and competition with deep reinforcement
  learning}.
\newblock {\em PLOS ONE} 12(4):e0172395.

\bibitem[\protect\citeauthoryear{Torrado \bgroup et al\mbox.\egroup
  }{2018}]{torrado2018deep}
Torrado, R.~R.; Bontrager, P.; Togelius, J.; Liu, J.; and Perez-Liebana, D.
\newblock 2018.
\newblock {Deep Reinforcement Learning for General Video Game AI}.
\newblock {\em arXiv preprint arXiv:1806.02448}.

\bibitem[\protect\citeauthoryear{Tuyls and Weiss}{2012}]{Tuyls:2012up}
Tuyls, K., and Weiss, G.
\newblock 2012.
\newblock {Multiagent learning: Basics, challenges, and prospects}.
\newblock {\em AI Magazine} 33(3):41--52.

\bibitem[\protect\citeauthoryear{Zahavy, Ben-Zrihem, and
  Mannor}{2016}]{zahavy2016graying}
Zahavy, T.; Ben-Zrihem, N.; and Mannor, S.
\newblock 2016.
\newblock {Graying the black box: Understanding DQNs}.
\newblock In {\em ICML}.

\bibitem[\protect\citeauthoryear{Zhao and Szafron}{2009}]{zhao2009learning}
Zhao, R., and Szafron, D.
\newblock 2009.
\newblock Learning character behaviors using agent modeling in games.
\newblock In {\em AIIDE}.

\bibitem[\protect\citeauthoryear{Zheng \bgroup et al\mbox.\egroup
  }{2018}]{deepbpr2018}
Zheng, Y.; Meng, Z.; Hao, J.; Zhang, Z.; Yang, T.; and Fan, C.
\newblock 2018.
\newblock A deep bayesian policy reuse approach against non-stationary agents.
\newblock In {\em NeurIPS}.
\newblock  962--972.

\end{thebibliography}
\bibliographystyle{aaai}

\end{document}